%% file: Article_v2.tex
\documentclass[amsmath,amssymb, nofootinbib,10pt,eqsecnum]{revtex4}
\usepackage{ graphicx, amsmath, amsmath, amssymb}

\input{defs}

\usepackage[breaklinks,  colorlinks, citecolor=blue]{hyperref}
\begin{document}

\title{$f\left(\mathcal{R}\right)$  gravity as a dark energy fluid}
\author{Richard A. Battye}
\email{richard.battye@manchester.ac.uk}
\affiliation{Jodrell Bank Centre for Astrophysics, School of Physics and Astronomy, The University of Manchester, Manchester M13 9PL, U.K.}
\author{Boris Bolliet}
\email{boris.bolliet@ens-lyon.fr}
\affiliation{Laboratoire de Physique Subatomique et de Cosmologie, Universit\'e Grenoble-Alpes, CNRS/IN2P3\\
53, avenue des Martyrs, 38026 Grenoble cedex, France}
\author{Jonathan A. Pearson}
\email{j.pearson@nottingham.ac.uk}
\affiliation{School of Physics and Astronomy, University of Nottingham, Nottingham NG7 2RD, U.K.}

\date{\today}

\begin{abstract} 
We study the equations for the evolution of cosmological perturbations in $f\left(\mathcal{R}\right)$ and conclude that this modified gravity model can be expressed as a dark energy fluid at background and linearised perturbation order. By eliminating the extra scalar degree of freedom known to be present in such theories, we are able to characterise the evolution of the perturbations in the scalar sector in terms of equations of state for the entropy perturbation and anisotropic stress which are written in terms of the density and velocity perturbations of the dark energy fluid and those in the matter, or the metric perturbations.  We also do the same in the much  simpler vector and tensor sectors. In order to illustrate the simplicity of this formulation, we numerically evolve perturbations in a small number of cases.
 \end{abstract}

\maketitle

\section{Introduction}

In the past few years there has been a growing realisation that  dark energy \cite{Copeland:2006wr} and modified gravity theories \cite{Clifton:2011jh} models need to be confronted with observational data in a systematic way. This has generated interest in constructing frameworks and formalisms for comparing models, or classes of models, to data as opposed to testing individual models. There are number of approaches which have been developed to do this including the Effective Field Theory for dark energy \cite{Bloomfield:2012ff, Gleyzes:2013ooa, Piazza:2013coa, Bloomfield:2013efa}, Parametrized Post Friedmann framework \cite{Skordis:2008vt, Baker:2011jy, Baker:2012zs, Ferreira:2014mja},  and the  Equation of State for perturbations (EoS) \cite{Battye:2012eu, Battye:2013aaa, Battye:2013ida}. These very similar ideas correspond to parameterizations at the level of the perturbed action, perturbed gravitational field equations, and the perturbed dark energy fluid equations, respectively. In this paper we will concentrate on the EoS approach.

Each of these parameterization schemes can be used in two different ways. The first is to construct arbitrary dark sector theories and the second is to map from a given model to the observationally combinations. In the latter of these ---``model-mapping'' --- EoS approach  prescribed how micro-physical degree of freedoms in the model combine to affect the evolution of quantities such as densities and velocity fields that are related to observables in terms of equations of state for the gauge invariant entropy and anisotropic stresses. Using the former approach preliminary constraints have been discussed in \cite{Soergel:2014sna,Battye:2014xna,Ade:2015rim} based on presently available cosmological data including that from the Cosmic Microwave Background (CMB), weak lensing and redshift space distortions (RSDs).

One of the most popular modified gravity theories is the $f(R)$ class of models \cite{Sotiriou:2008rp, DeFelice:2010aj}. The $f\left(\mathcal{R}\right)$ models of gravity are constructed by replacing the Ricci scalar in the Einstein-Hilbert action by an arbitrary function of the Ricci scalar $f\left(\mathcal{R}\right)$. Such models are well known to lead to an extra scalar degree of freedom and it has been shown, for example in \cite{2007PhRvD..76f4004H,2007PhLB..654....7A,2007JETPL..86..157S}, that observationally acceptable models can be constructed.

In this paper we provide expressions for the equations of state for perturbations which completely characterize the linearized perturbations in $f\left(\mathcal{R}\right)$ modified gravity, including the scalar, vector, and tensor modes.  Essentially we will model-map this theory into the EoS formalism. In doing this we show that the $f(\mathcal{R})$ modification to General Relativity (GR) can be formulated as a dark energy fluid at background order ---something which is well known ---  and also at first order in perturbations. As well as providing a physical interpretation and allowing these models to be included under the umbrella of the EoS formalism, we will see that writing the theory in this way can allow for a very simple inclusion in codes, such as {\tt CAMB}, used to calculate cosmological observables. 

\section{Background Field Equations}
\label{background}

The $f\left(\mathcal R\right)$ models  of gravity are characterized by the action
\be
\label{eq:FR-action-1}
S = \tfrac{1}{2}\int \dd^4x\, \sqrt{-g} \,  \left\{  {\mathcal{R} + f\left(\mathcal{R}\right)} \right\}  + \qsubrm{S}{m},
\ee
where $\mathcal{R}$ is the Ricci scalar and $\qsubrm{S}{m}$ is the action describing the standard matter fields. Natural units, $c=\hbar=\qsubrm{M}{Pl}=1$, are used throughout this paper. Varying the action (\ref{eq:FR-action-1}) with respect to the space-time metric $g_{\mu\nu}$ yields the field equations,
\bea
\label{eq:U-gen-fr-full}
G_{\mu\nu} =  T_{\mu\nu} + U_{\mu\nu},
\eea
where $G_{\mu\nu}$ is the Einstein tensor and $T_{\mu\nu}$ is the stress-energy tensor of the  standard matter fields. All contributions due to $f\left(\mathcal{R}\right)$  are packaged into the extra-term $U_{\mu\nu}$, which we call the stress-energy tensor of the dark sector, explicitly formulated as
\bea
\label{eq:U-gen-fr}
U_{\mu\nu} \defn  \tfrac{1}{2}fg_{\mu\nu}- \left( R_{\mu\nu}+g_{\mu\nu}\square - \nabla_{\mu}\nabla_{\nu}\right)f_{\mathcal{R}},
\eea
where $R_{\mu\nu}$ is the Ricci tensor, and $f_\mathcal{R}\equiv\frac{df}{d\mathcal{R}}$. Direct calculation shows that $U_{\mu\nu}$ is covariantly conserved, $\nabla^{\mu}U_{\mu\nu}=0$, as is required by the conservation of the matter energy-momentum tensor, $T_{\mu\nu}$. The background geometry is assumed to be isotropic and spatially flat, with a line element written as $ds^2=-dt^2+a\left(t\right)^2\delta_{ij}dx^{i} dx^{j}$, where $a\left(t\right)$ is the scale factor. Instead of the first and second order time derivative of the Hubble parameter, $H$, we use the dimensionless parameters\footnote{\label{fn:1}With these notation the Ricci scalar reads $\mathcal{R}=12H^2(1-\tfrac{1}{2}\qsubrm{\epsilon}{H})$. Furthermore,  $\qsubrm{\bar{\epsilon}}{H}$ and $\qsubrm{\epsilon}{H}$ are related through $\qsubrm{\bar{\epsilon}}{H}=\qsubrm{\epsilon}{H}^\prime+4\qsubrm{\epsilon}{H}-2\qsubrm{\epsilon}{H}^2$.}
\be
\qsubrm{\epsilon}{H}\equiv- \frac{H^\prime}{H},\,\,\,\,\,\,\,\,\,\,\,\,\,\,\qsubrm{\bar{\epsilon}}{H}\equiv-\frac{\mathcal{R}^\prime}{6H^2},
\ee
where the prime denotes derivative with respect to $d/d\ln a$. The dark sector can be viewed as a fluid, with energy density $\qsubrm{\rho}{de} \equiv \tfrac{1}{a^2} U_{00}$ and pressure $\qsubrm{P}{de} \equiv \tfrac{1}{3a^2} \delta^{ij}U_{ij}$. The field equations (\ref{eq:U-gen-fr-full}) can be recast as
\be
\qsubrm{\Omega}{m}+\qsubrm{\Omega}{de}=1,\,\,\,\,\,\,\,\,\,\,\,\,\,\,\qsubrm{w}{m}\qsubrm{\Omega}{m}+\qsubrm{w}{de}\qsubrm{\Omega}{de}=\tfrac{2}{3}\qsubrm{\epsilon}{H} -1,\label{eq:fe}
\ee
where $\qsubrm{\Omega}{i}=\frac{\qsubrm{\rho}{i}}{3H^{2}}$ and $\qsubrm{w}{i}\equiv \qsubrm{P}{i}/\qsubrm{\rho}{i}$ for $\mathrm{i}\in\left\{\mathrm{m,de}\right\}$. From (\ref{eq:U-gen-fr}), the density and equation of state parameters of the $f(\mathcal{R})$ fluid are explicitly given by
\bse
\begin{eqnarray}
\qsubrm{\Omega}{de} & = & -\frac{f}{6H^2}+(1-\qsubrm{\epsilon}{H})f_\mathcal{R}-f_\mathcal{R}^\prime,\label{eq:exrho}\\
\qsubrm{w}{de}+1  & = & -\frac{1}{3\qsubrm{\Omega}{de}}\left(2\qsubrm{\epsilon}{H}f_\mathcal{R}+(1+\qsubrm{\epsilon}{H})f_\mathcal{R}^\prime-f_\mathcal{R}^{\prime\prime}\right).\label{eq:exw}
\end{eqnarray}
\ese
The case of a cosmological constant is recovered  when $f(\mathcal{R})=-2\Lambda$, for which (\ref{eq:U-gen-fr-full}) reduces to the standard Einstein's field equations. Note that (\ref{eq:exrho}) is actually a second order differential equation for the function $f(\mathcal{R})$ since $f_\mathcal{R}=f^\prime/\mathcal{R}^\prime$. When the equation of state parameters, $\qsubrm{w}{i}$'s, are taken to be constant, this equation can be integrated leading to the so-called designer $f(\mathcal{R})$, see \cite{HSSS} for details.
\section{Gauge Invariant Formalism for Linear Perturbations}
\label{sec:gif}
The dynamics of linear perturbations is written in Fourier space, in both the synchronous and conformal Newtonian gauges. Instead of the coordinate wavenumber that appears in the Fourier transform, $k$, a reduced dimensionless wavenumber will be used, 
\bea
\mathrm{K}\equiv\frac{k}{aH},
\eea
so that $\mathrm{K}\ll 1$ and $\gg 1$ can be used identify the sub-(super)-horizon regimes. In the synchronous gauge, the non-zero metric perturbations are $\delta g_{ij}=a^2 h_{ij}$. In an orthonormal basis $\{\hat k, \hat l, \hat m\}$ in $k$-space, the spatial matrix $h_{ij}$ is further decomposed as 
$h_{ij}=\frac{1}{3}h\delta_{ij}+h_{\parallel}\sigma_{ij}+\qsuprm{h}{V}\cdot v_{ij}+\qsuprm{h}{T}\cdot e_{ij}$, where the notations $\qsuprm{h}{V(T)}$ contain the two vector (tensor) polarization states  (the dot product is to be understood as a sum over the polarization states).  Instead of $h$, we use the combination $6\eta\equiv h_{\parallel}-h$. The basis matrices are $\sigma_{ij} =\hat{k}_{i}\hat{k}_{j}-\frac{1}{3}\delta_{ij}$ for the longitudinal traceless mode, $\qsuprm{v}{(1)}_{ij}=2\hat{k}_{(i}\hat{l}_{j)}$ and $\qsuprm{v}{(2)}_{ij}=2\hat{k}_{(i}\hat{m}_{j)}$ for the vector modes and $e_{ij}^{\times}=2\hat{l}_{[i}\hat{m}_{j]}$, $e_{ij}^{+}=\hat{l}_{i}\hat{l}_{j}-\hat{m}_{i}\hat{m}_{j}$ for the tensor modes. In the conformal Newtonian gauge, the scalar modes are given by  $\delta{g_{00}}=-2a^{2}\psi$ and $\delta{g_{ij}}=-2a^{2}\phi\delta_{ij}$, while the tensor and vector modes remain the same in both gauges). An additional scalar degree of freedom arises at the perturbative level from the non-vanishing $f_\mathcal{R}^\prime$, given by
\be
\chi\equiv-\frac{f_\mathcal{R}^\prime}{\qsubrm{\bar{\epsilon}}{H}}\frac{\delta\mathcal{R}}{6H^2}.\label{eq:chidef}
\ee

This feature was pointed out in \cite{Bean:2006up} and is a manifestation of the well-known connection between $f(\mathcal R)$ theories and non-minimally coupled scalar-tensor theories \cite{Faulkner:2006ub,Sotiriou:2006hs}. Actually, $f_\mathcal{R}^\prime$ constitutes the first non-trivial contribution of an arbitrary function of the Ricci scalar since a linear or affine $f(\mathcal R)$ can always be recast as standard GR with a rescaled Newton constant \cite{HSSS}.

Our results are presented simultaneously in both the synchronous and conformal Newtonian gauges thanks  to a new set of variables presented below: $W,X,Y,Z$ for the GR sector and $\hat{\chi}, {\hat{\chi}}^\prime, {\hat{\chi}}^{\prime\prime}$ for the $f(\mathcal R)$ sector. The introduction of this set of variables is motivated by the gauge transformation rules that are recalled in Appendix \ref{append_gauge}. Quantities denoted with the subscript `S' (`C') are evaluated in the synchronous (conformal Newtonian) gauge.
\begin{eqnarray}
\begin{array}{ccc}
\quad \rbm{Symbol\quad} &\quad \rbm{Synchronous\,\,gauge\quad} & \quad\rbm{Conformal\,\,Newtonian\,\,gauge\quad}\\
T & \frac{{h}_{\parallel}^\prime}{2\rm{K}^{2}} & 0\\
Y & T^\prime+\qsubrm{\epsilon}{H}T & \psi\\
Z & \eta-T & \phi\\
X & Z^\prime+Y & Z^\prime+Y\label{eq:xdef}\\
W &X^\prime-\qsubrm{\epsilon}{H}(X+Y) & X^\prime-\qsubrm{\epsilon}{H}(X+Y)\\
\hat{\chi} & \chi_{s}+f_\mathcal{R}^\prime T &  \mathcal{\chi}_{c}\\
{\hat{\chi}}^\prime & \chi_{s}^\prime+(f_{\mathcal{R}}^{\prime\prime}-\qsubrm{\epsilon}{H}f_{\mathcal{R}}^{\prime})T & \chi_{c}^\prime-f_\mathcal{R}^\prime\psi\\
{\hat{\chi}}^{\prime\prime} & \chi_{s}^{\prime\prime}-\qsubrm{\epsilon}{H}\chi_{s}^\prime+(f_{\mathcal{R}}^{\prime\prime\prime}-3\qsubrm{\epsilon}{H}f_{\mathcal{R}}^{\prime\prime}+(4\qsubrm{\epsilon}{H}-\qsubrm{\bar{\epsilon}}{H})f_{\mathcal{R}}^{\prime})T & \chi_{c}^{\prime\prime}-\qsubrm{\epsilon}{H}\chi_{c}^\prime-f_\mathcal{R}^\prime\psi^\prime-2(f_{\mathcal{R}}^{\prime\prime}-\qsubrm{\epsilon}{H}f_{\mathcal{R}}^{\prime})\psi
\end{array}
\end{eqnarray}
Let us emphasize that ${\hat{\chi}}^\prime$ and ${\hat{\chi}}^{\prime\prime}$ are not just the first and second derivatives of $\hat{\chi}$, but are degrees of freedom in their own right. Using these gauge invariant variables, the first order perturbation of the Ricci scalar reads $\delta\mathcal{R}=-6H^2(W+4X-\tfrac{1}{3}K^2(Y-2Z)-\qsubrm{\bar{\epsilon}}{H}T)$. The fact that $T$ appears explicitly in the expression of $\delta\mathcal{R}$ indicates that this is not a gauge invariant quantity, and hence that $\chi$ defined in (\ref{eq:chidef}) is also not gauge invariant. However, $\hat{\chi}$  defined in the table above is gauge invariant and it can be written in terms of the geometric perturbations as
\begin{equation}
\hat{\chi}=\tfrac{f_{\mathcal{R}}^\prime}{\qsubrm{\bar{\epsilon}}{H}}\left\{W+4X-\tfrac{1}{3}\mathrm{K}^{2}(Y-2Z)\right\},\label{eq:hatchi}
\end{equation}
which is valid in both gauges, when $W,X,Y$ and $Z$ are replaced by the corresponding expressions presented in the table. 

A generic stress-energy tensor, ${D^{\mu}}_{\nu}$, can be decomposed into 
\begin{equation}
{\delta D^{\mu}}_{\nu}=\left(\rho\delta+\delta P\right)u^{\mu}u_{\nu}+\left(\rho+P\right)\left(u_{\nu}\delta u^{\mu}+u^{\mu}\delta u_{\nu}\right)+\delta P{\delta^{\mu}}_{\nu}+P{{\Pi}^{\mu}}_{\nu},
\label{eq:pset}
\end{equation}
where the density contrast is $\delta \equiv \delta \rho/\rho$, the Hubble flow is parametrized by $u_\nu=(-1,\vec{0})$ in coordinate time, and $\delta u_\nu=(0,\delta u_i)$ is the perturbed velocity field whose scalar mode is $\theta  \equiv  \frac{\ci k^{j}\delta u_{j}}{k^{2}}$. 

Instead of $\delta$ and $\theta$, we make an extensive use of the  dimensionless variables
\be
\Delta \equiv \delta+3\left(1+w\right)H\theta, \,\,\,\,\,\,\,\,\,\,\,\,\,\,\Theta \equiv 3\left(1+w\right)H\theta.
\ee
In the same way as for the geometric perturbations, it is possible to form gauge invariant combinations of the perturbed fluid variables:
\begin{eqnarray}
\begin{array}{ccc}
\quad \rbm{Symbol\quad} &\quad \rbm{Synchronous\,\,gauge\quad} & \quad\rbm{Conformal\,\,Newtonian\,\,gauge\quad}\\
\hat{\Theta} & \Theta_{s}+3\left(1+w\right)T & \Theta_{c}\\
\hat{\delta P} & \delta P_{s}+P_{s}^\prime T & \delta P_{c}
\end{array}
\end{eqnarray}
The gauge invariant pressure perturbation, $\hat{\delta P}$, is packaged into the gauge invariant entropy perturbation,
\begin{equation}
w\Gamma  =  \frac{\hat{\delta P}}{\rho}-\frac{dP}{d\rho}\left(\Delta-\hat{\Theta}\right)\label{eq:gammagi}.
\end{equation}
The anisotropic stress is the spatial traceless part of the stress-energy tensor.  In the same way as the metric perturbation, it decomposes into one scalar, $\pis$, two vector, $\qsuprm{\Pi}{V}$, and two tensor modes, $\qsuprm{\Pi}{T}$.  Note that our $\theta$ and $\qsuprm{\Pi}{S}$ differ from $\qsuprm{\theta}{MB}$ and $\sigma$ (anisotropic stress) defined in \cite{Ma:1995ey} by $\qsuprm{\theta}{MB}=\frac{k^2}{a}\theta$ and $(\rho+P)\sigma=-\frac{2}{3}P\qsuprm{\Pi}{S}$.

The generic perturbed fluid equations which follow from the conservation of the stress-energy tensor, $\delta(\nabla^\mu D_{\mu\nu})=0$, are 
\bse
\label{eq:pfe}
\begin{eqnarray}
{\Delta^\prime}-3w\Delta-2w\qsuprm{\Pi}{S}+\qsubrm{g}{K}\qsubrm{\epsilon}{H}\hat{\Theta} & = &3\left(1+w\right)X,\label{eq:deltadot}\\
{\hat{\Theta}}^{\prime}+3(\tfrac{dP}{d\rho}-w+\tfrac{1}{3}\qsubrm{\epsilon}{H})\hat{\Theta}-3\tfrac{dP}{d\rho}\Delta-2w\qsuprm{\Pi}{S}-3w\Gamma & = & 3\left(1+w\right)Y,\label{eq:Thetadot}
\end{eqnarray}
\ese
where
\be
\qsubrm{g}{K}\equiv1+\frac{\rm{K}^2}{3\qsubrm{\epsilon}{H}}. 
\ee

The field equations (\ref{eq:U-gen-fr-full}) expanded to linear order in perturbations, $\delta G_{\mu\nu}=\delta T_{\mu\nu} +\delta U_{\mu\nu}$, yield
\bse
\label{eq:dgdu}
\begin{eqnarray}
-\tfrac{2}{3}\mathrm{K}^{2}Z & = & \qsubrm{\Omega}{m}\qsubrm{\Delta}{m}+\qsubrm{\Omega}{de}\qsubrm{\Delta}{de},\label{eq:deltam}\\
2X & = & \qsubrm{\Omega}{m}\qsubrm{\hat{\Theta}}{m} + \qsubrm{\Omega}{de}\qsubrm{\hat{\Theta}}{de},\label{eq:Thetam}\\
\tfrac{2}{3}W+2X-\tfrac{2}{9}\mathrm{K}^{2}\left(Y-Z\right) & = &\qsubrm{\Omega}{m}(\qsubrm{\hat{\delta P}}{m}/\qsubrm{\rho}{m}) +  \qsubrm{\Omega}{de}(\qsubrm{\hat{\delta P}}{de}/\qsubrm{\rho}{de}),\label{eq:deltaP_m}\\
\tfrac{1}{3}\mathrm{K}^{2}\left(Y-Z\right) & = &\qsubrm{\Omega}{m}\qsubrm{w}{m}\qsubrm{\Pi}{m}^{\scriptscriptstyle{\mathrm{S}}} +  \qsubrm{\Omega}{de} \qsubrm{w}{de}\qsubrm{\Pi}{de}^{\scriptscriptstyle{\mathrm{S}}},\label{eq:Pi_Sm}\\
\tfrac{1}{6}{\qsuprm{h}{V}}^{\prime\prime}+(\tfrac{1}{2}-\tfrac{1}{6}\qsubrm{\epsilon}{H}){\qsuprm{h}{V}}^\prime & = &\qsubrm{\Omega}{m}\qsubrm{w}{m}\qsubrm{\Pi}{m}^{\scriptscriptstyle{\mathrm{V}}}+  \qsubrm{\Omega}{de} \qsubrm{w}{de}\qsubrm{\Pi}{de}^{\scriptscriptstyle{\mathrm{V}}},\label{eq:Pi_V_m}\\
\tfrac{1}{6}{\qsuprm{h}{T}}^{\prime\prime}+(\tfrac{1}{2}-\tfrac{1}{6}\qsubrm{\epsilon}{H}){\qsuprm{h}{T}}^\prime+\tfrac{1}{3}\mathrm{K}^{2}\qsuprm{h}{T} & = & \qsubrm{\Omega}{m}\qsubrm{w}{m}\qsubrm{\Pi}{m}^{\scriptscriptstyle{\mathrm{T}}} +  \qsubrm{\Omega}{de} \qsubrm{w}{de}\qsubrm{\Pi}{de}^{\scriptscriptstyle{\mathrm{T}}}.\label{eq:Pi_T_m}
\end{eqnarray}
\ese
The first equation (\ref{eq:deltam}) enables to write $\mathrm{K}^2Z$ in terms of the $\qsubrm{\Delta}{i}$'s, while the second equation (\ref{eq:Thetam}) constitutes the expression of the metric perturbation $X$ in terms of the perturbed fluid variables $\qsubrm{\hat{\Theta}}{i}$'s. The variables $\left\{W,X,Y,Z, \hat{\chi}, {\hat{\chi}}^\prime, {\hat{\chi}}^{\prime\prime}, \qsuprm{h}{V,T}\right\}$, which are linear combinations of the metric perturbations and their time derivatives, are called the \textit{geometric perturbations}. The variables $\left\{\qsubrm{\Delta}{i}, \qsubrm{\hat{\Theta}}{i}, \qsubrm{\delta \hat{P}}{i},  \qsubrm{\Gamma}{i}, \qsubrm{\Pi}{i}^{\scriptscriptstyle{\mathrm{S,V,T}}}\right\}$ with $\mathrm{i}\in\left\{\mathrm{m,de}\right\}$, which are linear combinations of the different projections of a perturbed stress-energy tensor, are called the \textit{perturbed fluid variables}. 
\section{Equation of State for Perturbations}
Equations of state for perturbations (EoS) constitute expressions for the entropy perturbation, $\qsubrm{\Gamma}{de}$, and the anisotropic stresses, $\qsubrm{\Pi}{de}^{\scriptscriptstyle{\mathrm{S,V,T}}}$,  that are constructed out of the perturbed metric degrees of freedom, the matter fluid variables, and  the dark density and velocity divergence fields. Once these expressions are provided, the equations governing cosmological perturbations explicitly closes. Schematically, for the scalar sector in synchronous gauge, we are looking to obtain expressions of the form
\bea
\qsubrm{\Gamma}{de} = \qsubrm{\Gamma}{de}  ( \qsubrm{\delta}{de}, \qsubrm{\theta}{de}, h^\prime, \eta,  \ldots, \qsubrm{\delta}{m}),\qquad
\qsubrm{\Pi}{de}^{\scriptscriptstyle{\mathrm{S}}} = \qsubrm{\Pi}{de}^{\scriptscriptstyle{\mathrm{S}}}  ( \qsubrm{\delta}{de}, \qsubrm{\theta}{de}, h^\prime, \eta, \ldots, \qsubrm{\delta}{m} ),
\eea
where the list of arguments shown is not exhaustive and can include derivatives, for example. Certain classes of equation of state have already been worked out (see \cite{Battye:2013aaa, Battye:2013ida} for kinetic gravity braiding models, \cite{Bloomfield:2013cyf} for coupled Horndeski theories,  \cite{Gleyzes:2014rba} for generalised scalar-tensor theories and \cite{Battye:2007aa,2013PhRvD..88h4004B,Pearson:2014iaa} for relativistic elastic and viscoelastic material models).

The simplest way to understand this approach is in the vector and tensor sectors of the theory. If one assumes that there are no extra vector and tensor degrees of freedom (which is the case in $f(\mathcal{R})$ theories), then the anisotropic stresses can only be functions of the metric variables. Focusing on tensor modes, the only tensor field available is  the tensor mode of the metric perturbation, $\qsuprm{h}{T}$, and its time derivatives (which we will limit to second order). The most general form of $\qsubrm{\Pi}{de}^{\scriptscriptstyle{\mathrm{T}}}$ would then be given by
\bea
\label{eq:tensor}
\qsubrm{\Pi}{de}^{\scriptscriptstyle{\mathrm{T}}} = {\mathcal{T}_1} {\qsuprm{h}{T}}^{\prime\prime}+ {\mathcal{T}_2} {\qsuprm{h}{T}}^{\prime}+{\mathcal{T}_3}\qsuprm{h}{T},
\eea
where the $\{\mathcal{T}_i\}$ are a set of dimensionless functions of space and time, that do not depend on the perturbed field quantities. Often one can deduce that these can be limited to just being functions of time only for specific theories. A similar expression could be written for the vector sector.

In $f({\mathcal{R}})$ gravity, the expansion to first order in perturbations of the dark sector stress-energy tensor is 
\begin{eqnarray}
\delta U_{\mu\nu} & = & -f_{\mathcal{R}}\delta R_{\mu\nu}+\tfrac{1}{2}f\delta g_{\mu\nu}+\tfrac{1}{2}g_{\mu\nu}f_{\mathcal{R}}\delta\mathcal{R}-f_{\mathcal{RR}}R_{\mu\nu}\delta\mathcal{R}\nonumber \\
 &  & \quad +\delta\left(\nabla_{\mu}\nabla_{\nu}f_{\mathcal{R}}\right)-\left(\square f_{\mathcal{R}}\right)\delta g_{\mu\nu}-g_{\mu\nu}\delta\left(\square f_{\mathcal{R}}\right).\label{eq:pee}
\end{eqnarray}
This allows us to isolate the perturbed fluid variables for the $f(\mathcal{R})$ dark sector theory \cite{Hwang}. The tensor and vector projections of (\ref{eq:pee}) readily constitute the EoS for $\qsubrm{\Pi}{de}^{\scriptscriptstyle{\mathrm{V}}}$ and $\qsubrm{\Pi}{de}^{\scriptscriptstyle{\mathrm{T}}}$,
\bse
\label{eq:dudefvt}
\begin{eqnarray}
\qsubrm{\Omega}{de} \qsubrm{w}{de}\qsubrm{\Pi}{de}^{\scriptscriptstyle{\mathrm{V}}} & = & -\tfrac{1}{6}f_{\mathcal{R}} {\qsuprm{h}{V}}^{\prime\prime}-\tfrac{1}{6}\left\{(3-\qsubrm{\epsilon}{H})f_{\mathcal{R}}+f_{\mathcal{R}}^\prime\right\}{\qsuprm{h}{V}}^\prime ,\label{eq:Pi_V}\\
 \qsubrm{\Omega}{de} \qsubrm{w}{de}\qsubrm{\Pi}{de}^{\scriptscriptstyle{\mathrm{T}}}& = & -\tfrac{1}{6}f_{\mathcal{R}} {\qsuprm{h}{T}}^{\prime\prime}-\tfrac{1}{6}\left\{(3-\qsubrm{\epsilon}{H})f_{\mathcal{R}}+f_{\mathcal{R}}^\prime\right\}{\qsuprm{h}{T}}^\prime-\tfrac{1}{6}f_{\mathcal{R}}\mathrm{K}^{2}\qsuprm{h}{T}. \label{eq:Pi_T}
\end{eqnarray}
\ese
As expected these are of the form (\ref{eq:tensor}) and the coefficients are just functions of time except for the explicit dependence on $\mathrm{K}^2$ in the final term in the expression for $\qsubrm{\Pi}{de}^{\scriptscriptstyle{\mathrm{T}}}$.

The scalar projections yield the following expressions:
\bse
\label{eq:dudef}
\begin{eqnarray}
\qsubrm{\Omega}{de}\qsubrm{\Delta}{de} & = & -\qsubrm{g}{K}\qsubrm{\epsilon}{H}\hat{\chi}+f_{\mathcal{R}}^\prime X+\tfrac{2}{3}f_{\mathcal{R}}\mathrm{K}^{2}Z,\label{eq:delta}\\
\qsubrm{\Omega}{de}\qsubrm{\hat{\Theta}}{de} & = & {\hat{\chi}}^\prime-\hat{\chi}-2f_{\mathcal{R}}X,\label{eq:Theta}\\
\qsubrm{\Omega}{de}(\qsubrm{\hat{\delta P}}{de}/\qsubrm{\rho}{de}) & = & \tfrac{1}{3}{\hat{\chi}}^{\prime\prime}+(\tfrac{2}{3}-\tfrac{1}{3}\qsubrm{\epsilon}{H}){\hat{\chi}}^\prime-\left(1-\tfrac{1}{3}\qsubrm{\epsilon}{H}-\tfrac{2}{9}\mathrm{K}^{2}\right)\hat{\chi}\nonumber\\
 &  & \qquad -\tfrac{2}{3}f_{\mathcal{R}}W-2(f_{\mathcal{R}}+\tfrac{1}{3}f_{\mathcal{R}}^\prime)X+\tfrac{2}{9}f_{\mathcal{R}}\mathrm{K}^{2}\left(Y-Z\right),\label{eq:deltaP}  \\
 \qsubrm{\Omega}{de} \qsubrm{w}{de}\qsubrm{\Pi}{de}^{\scriptscriptstyle{\mathrm{S}}} & = & -\tfrac{1}{3}\mathrm{K}^{2}\hat{\chi}-\tfrac{1}{3}f_{\mathcal{R}}\mathrm{K}^{2}\left(Y-Z\right).\label{eq:Pi_S}
\end{eqnarray}
\ese
From now on, the standard matter fluid will be assumed to have vanishing anisotropic stress and entropy perturbation, $\qsubrm{\Pi}{m}^{\scriptscriptstyle{\mathrm{S}}}=\qsubrm{\Gamma}{m}=0$ which is the case for a CDM fluid. When those are they are non-zero, the procedure presented below is easily generalized, with additional terms proportional to $\qsubrm{\Pi}{m}^{\scriptscriptstyle{\mathrm{S}}}$ and $\qsubrm{\Gamma}{m}$. 
In the last equation (\ref{eq:Pi_S}), $\qsubrm{\Pi}{de}^{\scriptscriptstyle{\mathrm{S}}}$ can be eliminated with (\ref{eq:Pi_Sm}), providing the expression of $Y$ in terms of $Z$ and $\hat{\chi}$,
\be
Y=Z-\tfrac{1}{1+f_\mathcal{R}}\hat{\chi},\label{eq:Y}
\ee 
valid for all $\mathrm{K}$. Therefore, the dark sector anisotropic stress is simply 
\be
 \qsubrm{\Omega}{de} \qsubrm{w}{de}\qsubrm{\Pi}{de}^{\scriptscriptstyle{\mathrm{S}}}=\tfrac{1-\qsubrm{g}{K}}{1+f_\mathcal{R}}\qsubrm{\epsilon}{H}\hat{\chi}.\label{eq:pisde}
\ee
Equation (\ref{eq:delta}), combined to (\ref{eq:deltam}) and (\ref{eq:Thetam}), enables to write $\hat{\chi}$ in terms of the $\qsubrm{\Delta}{i}$'s and $\qsubrm{\hat{\Theta}}{i}$'s,
\be
\hat{\chi}  = -\tfrac{\qsubrm{\Omega}{de}}{\qsubrm{g}{K}\qsubrm{\epsilon}{H}}\qsubrm{\Delta}{de} -\tfrac{f_{\mathcal{R}}}{\qsubrm{g}{K}\qsubrm{\epsilon}{H}}\left\{\qsubrm{\Omega}{de}(\qsubrm{\Delta}{de}-\tfrac{f_{\mathcal{R}}^\prime}{2f_{\mathcal{R}}} \qsubrm{\hat{\Theta}}{de})+\qsubrm{\Omega}{m}(\qsubrm{\Delta}{m}-\tfrac{f_{\mathcal{R}}^\prime}{2f_{\mathcal{R}}} \qsubrm{\hat{\Theta}}{m})\right\}.\label{eq:chi}
\ee
With (\ref{eq:pisde}, \ref{eq:chi}) one obtains the EoS for the anisotropic stress: 
\be
\qsubrm{w}{de}\qsubrm{\Pi}{de}^\mathrm{S}  = \tfrac{1}{3\qsubrm{g}{K}\qsubrm{\epsilon}{H}}\mathrm{K}^2\left\{\qsubrm{\Delta}{de} -\tfrac{f_{\mathcal{R}}^\prime}{2(1+f_\mathcal{R})} \qsubrm{\hat{\Theta}}{de}+\tfrac{\qsubrm{\Omega}{m}}{\qsubrm{\Omega}{de}}\tfrac{f_{\mathcal{R}}}{1+f_\mathcal{R}}\qsubrm{\Delta}{m}-\tfrac{\qsubrm{\Omega}{m}}{\qsubrm{\Omega}{de}}\tfrac{f_{\mathcal{R}}^\prime}{2(1+f_\mathcal{R})} \qsubrm{\hat{\Theta}}{m}\right\}.\label{eq:eosPi_S}
\ee

In order to deduce the entropy perturbation as an equation of state, one might begin with the expression for $\qsubrm{\hat{\delta P}}{de}$ in (\ref{eq:deltaP}), and then eliminate $\hat\chi$ and  its time-derivatives, as in \cite{Battye:2013aaa}. This involved  differentiation of (\ref{eq:delta}) or (\ref{eq:Theta}) to obtain $\hat{\chi}^{\prime\prime}$, and $\hat{\chi}^\prime$. However, this strategy  eventually leads to the perturbed fluid equation (\ref{eq:Thetadot}) and hence a tautology, and therefore an alternative strategy is required. The starting point is the field equation (\ref{eq:deltaP_m}). On the right-hand-side of (\ref{eq:deltaP_m}), the pressure perturbations $\qsubrm{\hat{\delta P}}{de}$ and  $\qsubrm{\hat{\delta P}}{m}$ are replaced in favor of the $\qsubrm{\Gamma}{i}$'s with (\ref{eq:gammagi}). On the left-hand-side of  (\ref{eq:deltaP_m}), $W$ is replaced in terms of the geometric perturbations $X,Y,Z$ and $\hat\chi$ with (\ref{eq:hatchi}). As before, $X$ is written in terms of the $\qsubrm{\hat{\Theta}}{i}$'s with (\ref{eq:Thetam}). Furthermore, with (\ref{eq:Y}) and  (\ref{eq:chi}), $Y$ can be expressed with $Z$ and the perturbed fluid variables $\qsubrm{\Delta}{i}$'s and $\qsubrm{\hat{\Theta}}{i}$'s. After these replacements, the geometric perturbations only appear within $\mathrm{K}^2 Z$ which can be replaced with the $\qsubrm{\Delta}{i}$'s with (\ref{eq:deltam}). Eventually, the EoS for the dark sector entropy perturbation is obtained as
\bea
\qsubrm{w}{de}\qsubrm{\Gamma}{de}& = &\left[\qsubrm{\zeta}{de}-\tfrac{\qsubrm{\bar{\epsilon}}{H}}{3\qsubrm{g}{K}\qsubrm{\epsilon}{H}}\tfrac{2(1+f_\mathcal{R})-f_\mathcal{R}^\prime}{f_\mathcal{R}^\prime}\right]\qsubrm{\Delta}{de}-\qsubrm{\zeta}{de}\qsubrm{\hat{\Theta}}{de}\nonumber\\
& &+\tfrac{\qsubrm{\Omega}{m}}{\qsubrm{\Omega}{de}}\left[\qsubrm{\zeta}{m}-\tfrac{\qsubrm{\bar{\epsilon}}{H}}{3\qsubrm{g}{K}\qsubrm{\epsilon}{H}}\tfrac{2f_\mathcal{R}-f_\mathcal{R}^\prime}{f_\mathcal{R}^\prime}\right]\qsubrm{\Delta}{m}-\tfrac{\qsubrm{\Omega}{m}}{\qsubrm{\Omega}{de}}\qsubrm{\zeta}{m}\qsubrm{\hat{\Theta}}{m}\label{eq:eosGamma}
\eea
where
\be
\qsubrm{\zeta}{i} \equiv\frac{\qsubrm{g}{K}\qsubrm{\epsilon}{H}-\qsubrm{\bar{\epsilon}}{H}}{3\qsubrm{g}{K}\qsubrm{\epsilon}{H}}-\frac{\qsubrm{d{P}}{i}}{\qsubrm{d{\rho}}{i}}. 
\ee
Note that when the matter fluid is pressure-less and $\qsubrm{w}{de}=-1$, then $\qsubrm{\bar{\epsilon}}{H}=\qsubrm{\epsilon}{H}$ and therefore
\be
\qsubrm{\zeta}{de}=\tfrac{4\qsubrm{g}{K}-1}{3\qsubrm{g}{K}},\,\,\,\,\,\,\,\,\,\,\,\,\,\,\,\,\,\,\qsubrm{\zeta}{m}=\tfrac{\qsubrm{g}{K}-1}{3\qsubrm{g}{K}}.
\ee

Equations (\ref{eq:eosPi_S}) and (\ref{eq:eosGamma}), as well as (\ref{eq:Pi_V}, \ref{eq:Pi_T}) for the vector and tensor sectors,  are the main result of this paper. They constitute the EoS for perturbations in $f(R)$ gravity expressed in a gauge invariant way.  One could choose to express the EoS in terms of the dark sector perturbed fluid variables and the geometric perturbations $X,Y$ and $Z$. For the entropy perturbation, this is achieved by replacing $\qsubrm{\Delta}{m}$ and $\qsubrm{\hat{\Theta}}{m}$ in (\ref{eq:eosGamma}) with (\ref{eq:deltam}) and (\ref{eq:Thetam}). The dark sector  EoS for perturbations are then expressed in a `self-consistent' way, which does not depend explicitly on the perturbed fluid variables of the other fluid components
\bse
\label{eq:eosFL}
\bea
\qsubrm{w}{de}\qsubrm{\Pi}{de}^\mathrm{S}&=&\tfrac{1}{3\qsubrm{\Omega}{de}}\mathrm{K}^2(Y-Z),\label{eq:eosPi_SDE}\\
\qsubrm{w}{de}\qsubrm{\Gamma}{de} &=& -\tfrac{\qsubrm{dP}{de}}{\qsubrm{d\rho}{de}}(\qsubrm{\Delta}{de}-\qsubrm{\hat{\Theta}}{de})+\tfrac{2}{3}\tfrac{\qsubrm{\bar{\epsilon}}{H}}{\qsubrm{\Omega}{de}}\tfrac{1+f_\mathcal{R}}{f_\mathcal{R}^\prime}(Z-Y)-\tfrac{2}{3\qsubrm{\Omega}{de}}(X+\tfrac{1}{3}\mathrm{K}^2 Z)\label{eq:eosGammaDE}.
\eea
\ese

\begin{figure}
\begin{center}
	\includegraphics[scale=0.5]{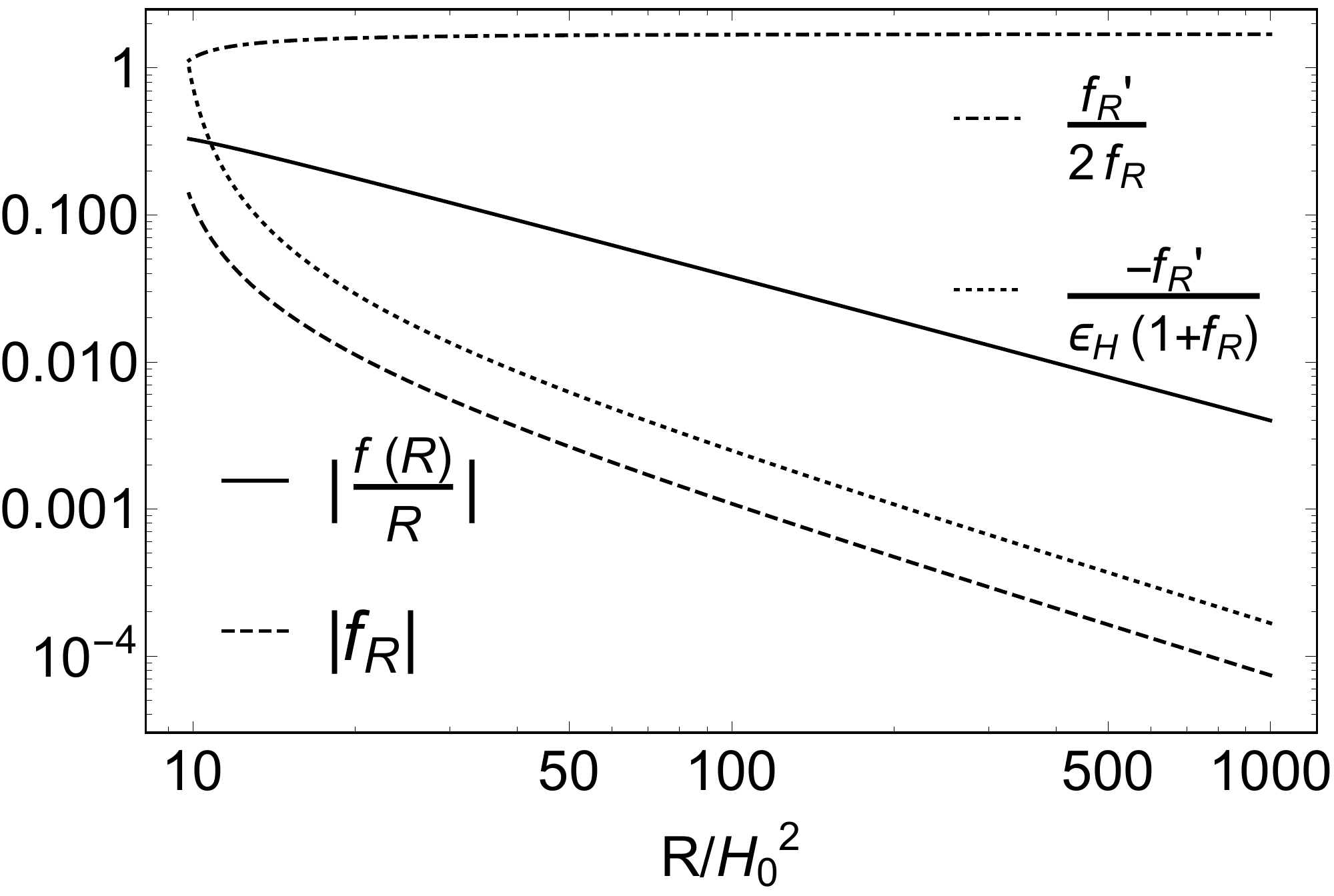}
\caption{Evolution of some relevant $f(\mathcal{R})$ quantities that appear as coefficients in the EoS for the dark sector anisotropic stress (\ref{eq:eosPi_S}) and entropy perturbation (\ref{eq:eosGamma}), for a designer $f(\mathcal{R})$ that mimics $\Lambda$-CDM ($\qsubrm{w}{de}=-1$) with $\qsubrm{B}{0}=1$. When these are negative valued, their absolute value is plotted in order to accommodate the logarithmic scale.}
\label{fig:figuref}
\end{center}
\end{figure}

The coefficients that play an important role in the EoS are either proportional to $f_\mathcal{R}$, $\frac{2f_\mathcal{R}}{f_\mathcal{R}^\prime}$ and
\be
{B}\equiv-\tfrac{f_\mathcal{R}^\prime}{\qsubrm{\epsilon}{H}(1+f_\mathcal{R})},\label{eq:bdef}
\ee
or its inverse. Their evolution in the case of a designer  $f(\mathcal R)$ with $\qsubrm{w}{de}=-1$ is plotted in figure \ref{fig:figuref}. In order to illustrate some of the applications of the EoS formalism and the gauge invariant notations, we shall now describe the procedure for solving the linear perturbations in $f(\mathcal R)$ gravity. 

\section{Dynamics of linear perturbation in $f(\mathcal R)$ gravity}
The dynamics of vector and tensor perturbations is straightforward to deduce and therefore we will only focus on the scalar sector. The dynamics of the scalar perturbations can be specified by writing the four perturbed fluid equations (\ref{eq:pfe}), plus one evolution equation for the geometric perturbation $Z$ which follows from the definition of the gauge invariant notations (\ref{eq:xdef}),
\be
Z^\prime=X-Y, \label{eq:zprime}
\ee
where $X$ and $Y$ are given in (\ref{eq:Thetam}, \ref{eq:Y}-\ref{eq:chi}) in terms of the perturbed fluid variables. In fact, the system of five differential equations is overdetermined: when $\mathrm{K}=0$, the field equation (\ref{eq:deltam}) can be used to express $\qsubrm{\Delta}{de}$ in terms of $\qsubrm{\Delta}{m}$; and when $\mathrm{K}\neq0$, the same equation gives $Z$ in terms of the  $\qsubrm{\Delta}{i}$'s. This approach is powerful and elegant: it provides an efficient way to solve the linear perturbation in $f(R)$ gravity, and the phenomenology becomes transparent through the interpretation of the fluid variables. 

\begin{figure}
\begin{center}
	\includegraphics[scale=0.4]{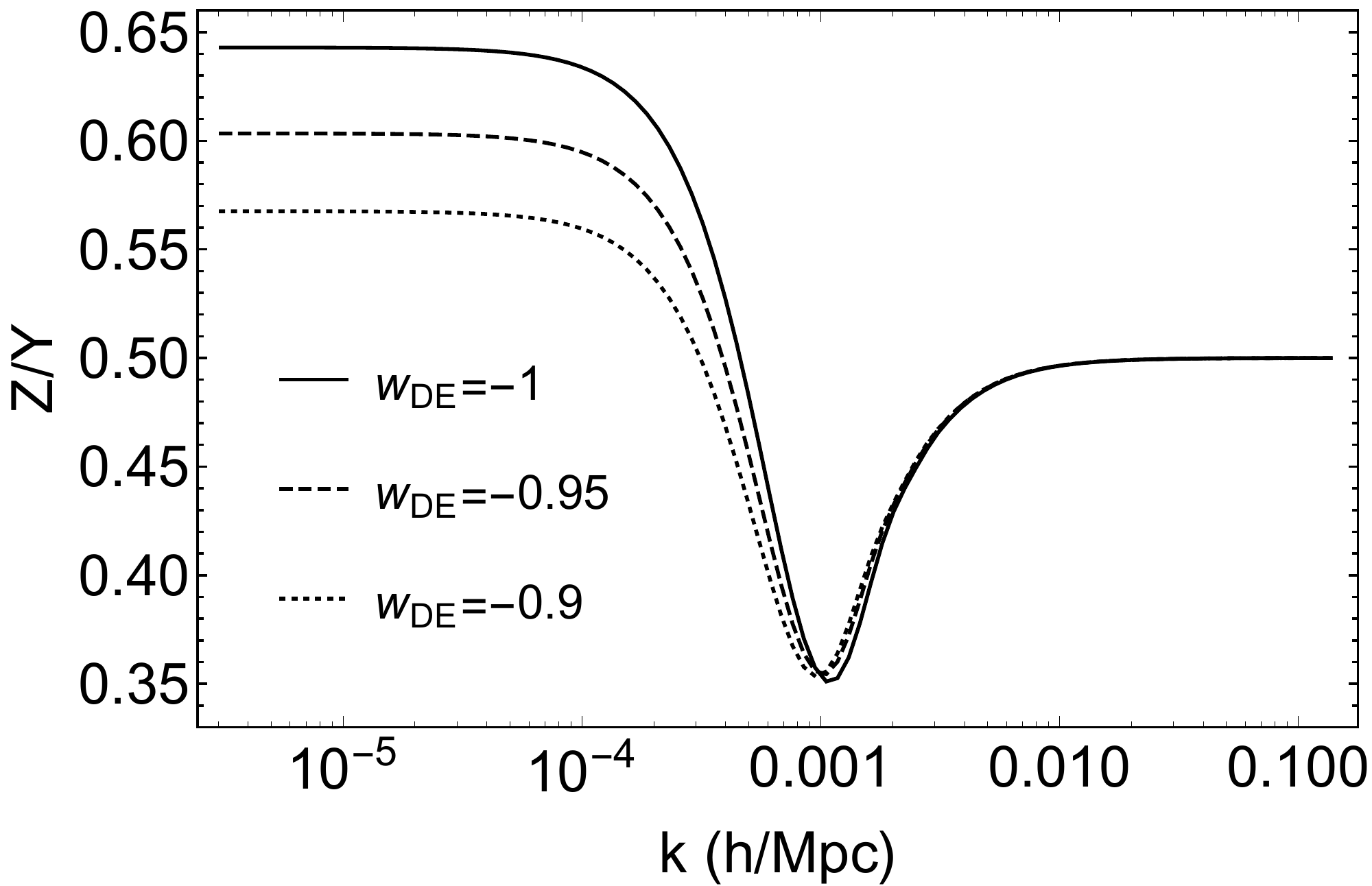} \includegraphics[scale=0.39]{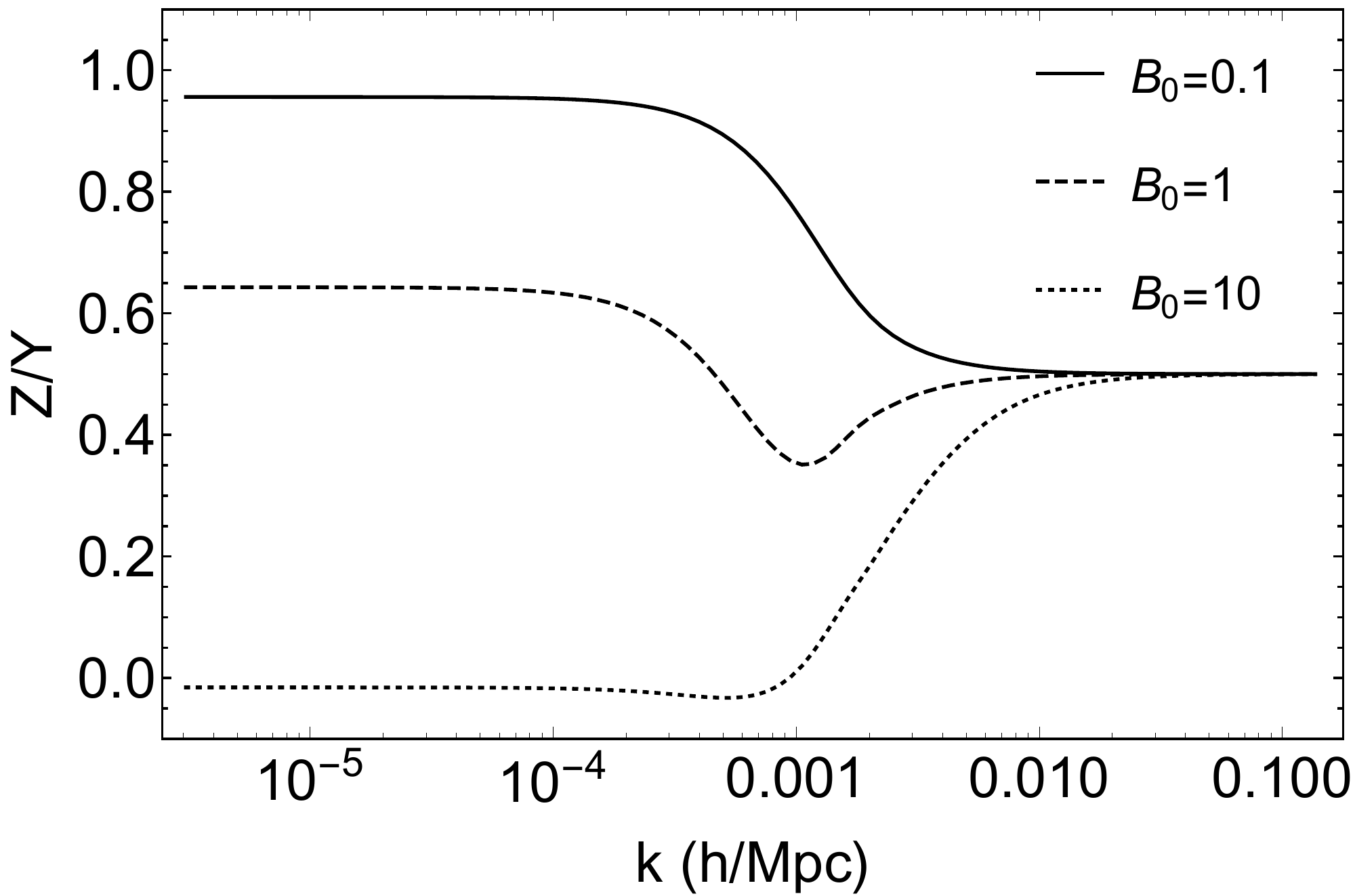}
\caption{Spectrum of the ratio $Z/Y$ (or $-\Phi/\Psi$ in the notation of \cite{HSSS}) for different values of the equation of state parameter when $\qsubrm{B}{0}=1$ (left) and different designer $f(\mathcal R)$ scenarios parametrized by $\qsubrm{B}{0}$ and with $\qsubrm{w}{de}=-1$ (right). On the x-axis, the wavenumber is written in units `$h/\mathrm{Mpc}$', where $h=0.73$ is the reduced Hubble constant.}
\label{fig:figureZY}
\end{center}
\end{figure}

Before proceeding to this analysis, let us note that an essential feature of  linear perturbation in $f(\mathcal{R})$ gravity can be deduced from (\ref{eq:Y}), when $\hat{\chi}$ is replaced by its expression (\ref{eq:hatchi}) in terms of the geometric perturbations,
\be
B\mathrm{K^2}(Y-2Z)=3BW+12BX+3\tfrac{\qsubrm{\bar{\epsilon}}{H}}{\qsubrm{\epsilon}{H}}(Y-Z),
\ee
where $B$ is defined in (\ref{eq:bdef}). Since the geometric perturbations $\left\{W,X,Y,Z\right\}$ shall remain bounded during their evolution, it appears that for $\mathrm{K}$ larger than $B^{-1/2}$ the ratio $Z/Y$ is driven to $1/2$, as illustrated in figure \ref{fig:figureZY} in the case of a designer $f(\mathcal{R})$.

\begin{figure}
\begin{center}
	\includegraphics[scale=0.4]{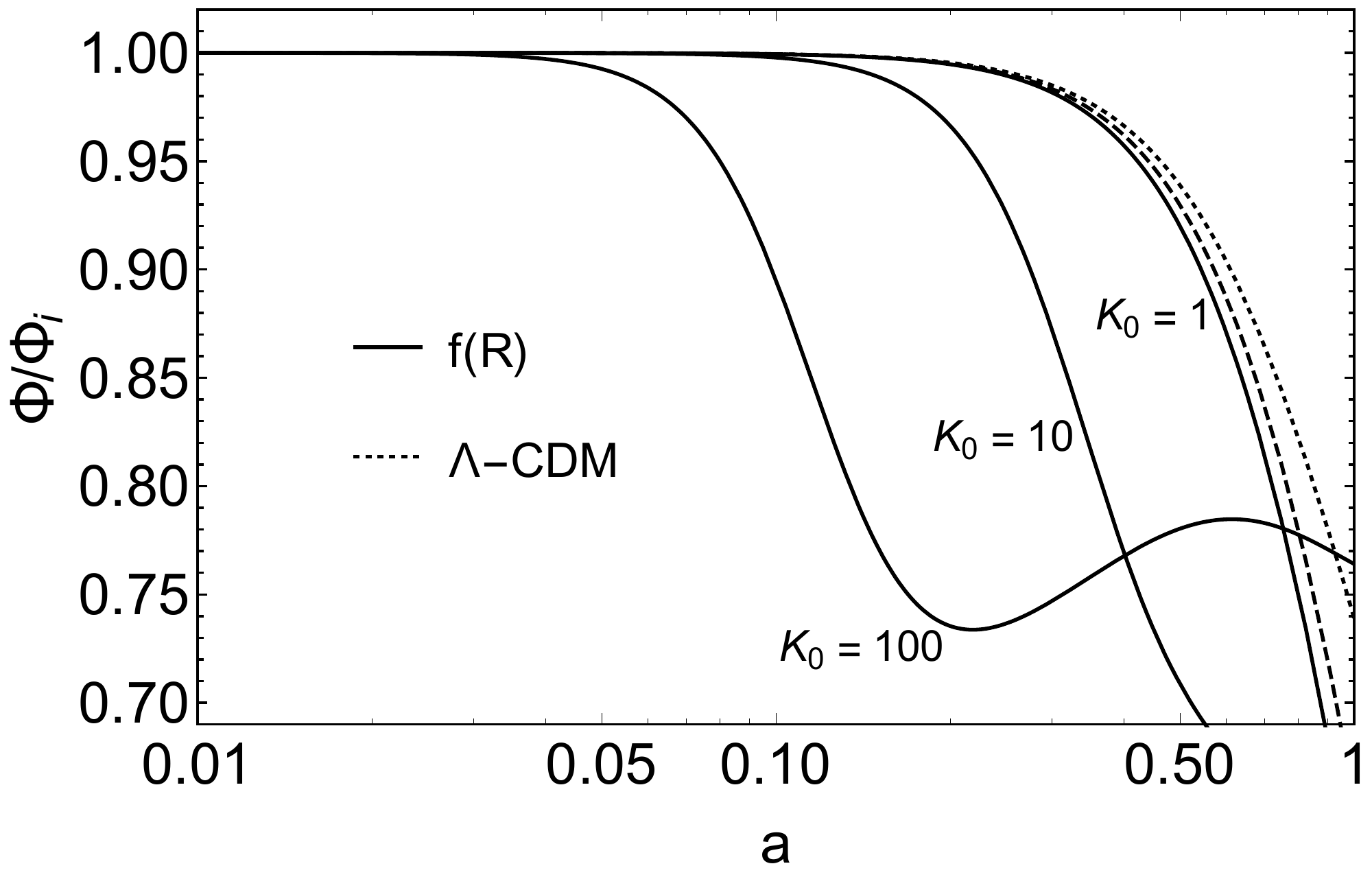} \includegraphics[scale=0.39]{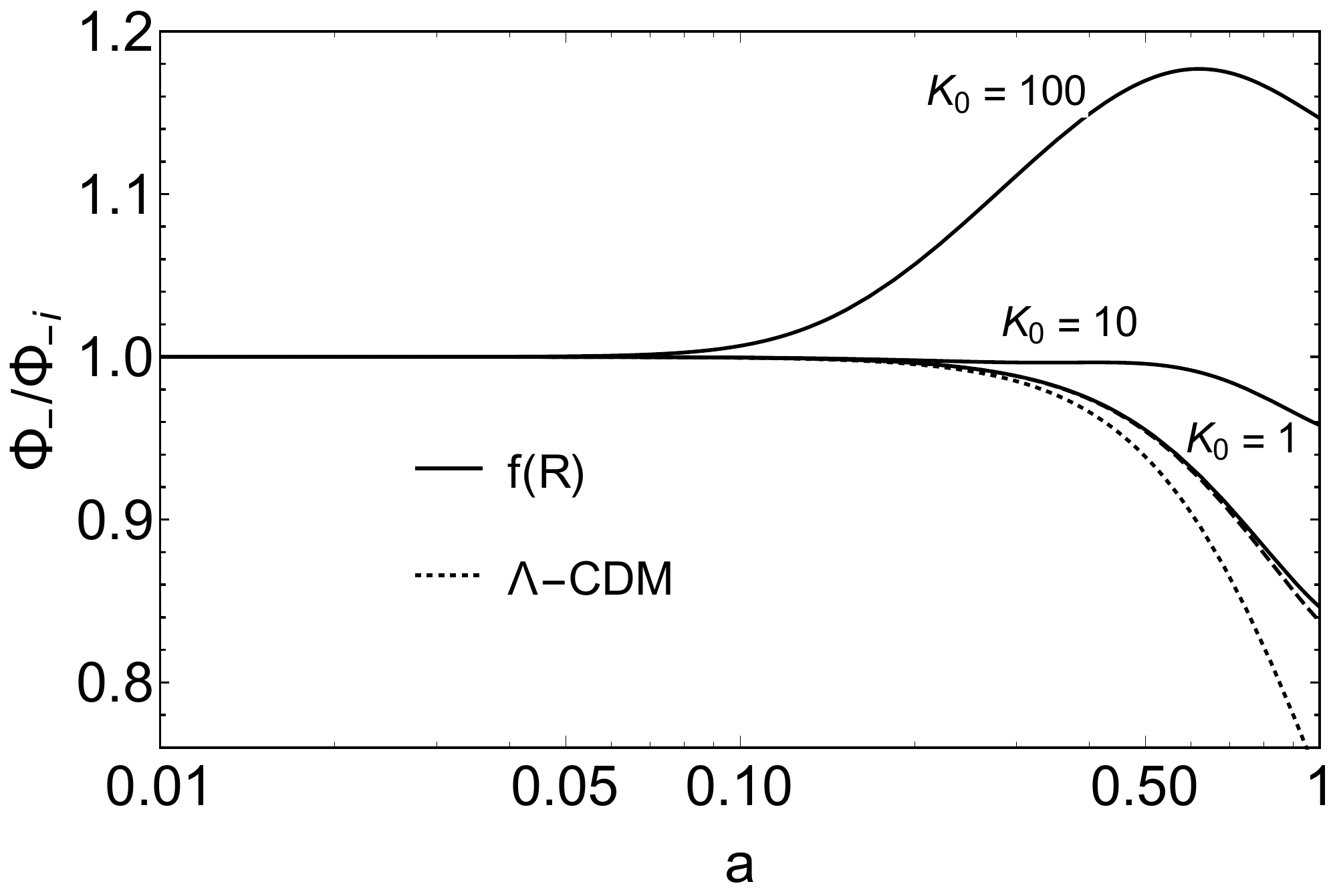}
\caption{Evolution of the metric perturbations $\Phi=-Z$ (left) and $\Phi_{-}=-\frac{Y+Z}{2}$ (right) for the $\Lambda$CDM expansion history ($\qsubrm{w}{de}= -1$, $\qsubrm{\Omega}{de}^0=0.76$). The dotted lines correspond to the $\Lambda$CDM scenario without $f(\mathcal R)$. In this case the amplitudes of the perturbed potentials $\Phi$ and $\Phi_{-}$ do not depend on the wavenumber. The black lines correspond to a $f(\mathcal R)$ scenario that mimics the cosmological constant ($\qsubrm{w}{de}= -1$, $\qsubrm{B}{0}=1$). The amplitudes of the metric perturbations are now sensitive to the wavenumber $\qsubrm{\mathrm{K}}{0}=\frac{k}{\qsubrm{H}{0}}$. Even the infrared limit, $\qsubrm{K}{0}=0$, represented by the dashed curves closest to the dotted lines, there is a disagreement with the $\Lambda$CDM predictions.}
\label{fig:figure2}
\end{center}
\end{figure}

Let us consider the case of a matter fluid with $\qsubrm{w}{m}=\qsubrm{\Pi}{m}^\mathrm{S}=\qsubrm{\Gamma}{m}=0$, along with a designer $f(\mathcal R)$ fluid that mimics $\Lambda$-CDM ($\qsubrm{w}{de}=-1$). The  function $f(\mathcal R)$ is determined by (\ref{eq:exrho}). As shown in  \cite{HSSS}, the different solutions to (\ref{eq:exrho}) can by parametrized by the single number $B$, defined in (\ref{eq:bdef}), evaluated today\footnote{A subscript `0' means that the quantity is evaluated today.} when $\qsubrm{a}{0}=1$ (analytical expressions for $f(\mathcal {R})$ are available  in some regimes \cite{He:2012rf}). For the numerical simulation we have chosen $\qsubrm{B}{0}=1$, and we have set the initial conditions for the perturbations at redshift $z=100$, when $B\ll1$. At such high curvature, during the matter dominated era, the initial conditions for the perturbations must follow from the general relativistic expectation in order to be consistent with CMB observations. Hence, initially ${\qsubrm{\Delta}{de}}={\qsubrm{\hat{\Theta}}{de}} =0$ and $\qsubrm{\Omega}{m}{\qsubrm{\Delta}{m}} =-\tfrac{2}{3}\mathrm{K^2}Z$, $\qsubrm{\Omega}{m}{\qsubrm{\hat{\Theta}}{m}}=2X$, with $X=Y=Z$. As mentioned before, the five relevant dynamical equations are

\begin{eqnarray}
\label{eq:pfemde}
\begin{array}{lllllll}
{\qsubrm{\Delta^\prime}{de}} & = &-3\qsubrm{\Delta}{de}-\qsubrm{g}{K}\qsubrm{\epsilon}{H}\qsubrm{\hat{\Theta}}{de}-2\qsubrm{\Pi}{de}^\mathrm{S}, & \quad\quad& {\qsubrm{\Delta^\prime}{m}} & = & -\qsubrm{g}{K}\qsubrm{\epsilon}{H}\qsubrm{\hat{\Theta}}{m} +3X,\\
{\qsubrm{\hat{\Theta}}{de}}^{\prime}& = &-3\qsubrm{\Delta}{de}-\qsubrm{\epsilon}{H}\qsubrm{\hat{\Theta}}{de}-2\qsubrm{\Pi}{de}^\mathrm{S}-3\qsubrm{\Gamma}{de}, & \quad\quad& {\qsubrm{\hat{\Theta}}{m}}^{\prime} & = &-\qsubrm{\epsilon}{H}\qsubrm{\hat{\Theta}}{m}+ 3Y,\\
Z^\prime& = &X-Y, & & & &
\end{array}
\end{eqnarray}
where $X$ and $Y$ are replaced with (\ref{eq:Thetam},  \ref{eq:Y}-\ref{eq:chi}) in terms of the perturbed fluid variables, while $\qsubrm{\Pi}{de}^\mathrm{S}$ and $\qsubrm{\Gamma}{de}$ are given in (\ref{eq:eosPi_S}) and (\ref{eq:eosGamma}). Recall that in the conformal Newtonian gauge, $Z=\phi$ and $Y=\psi$ in our notations that follow \cite{Ma:1995ey}, while $Z=-\Phi$ and $Y=\Psi$ in Song-Hu-Sawicki notation \cite{HSSS}. With this strategy we have successfully reproduced the results presented in figure 2 of \cite{HSSS}, see figure \ref{fig:figure2}.

Alternatively one could have favored the geometric perturbations instead of the perturbed fluid variables. The way the equations have been written makes it straightforward to go from one picture to the other. Let us pick the conformal Newtonian gauge in order to illustrate this point. In the conformal Newtonian gauge, $\hat{\chi}^\prime=\chi_{c}^\prime-f_\mathcal{R}^\prime Y$ (see section \ref{sec:gif}). Therefore, equation (\ref{eq:Theta}) combined to (\ref{eq:Y}) and (\ref{eq:zprime}) yields a first order differential equation for the geometric perturbation $Y$. The differential equation for $X$ is simply provided by (\ref{eq:hatchi}), where $\hat{\chi}$ is replaced with (\ref{eq:Y}) and $W$ with its definition, $W=X^\prime-\qsubrm{\epsilon}{H}(X+Y)$. Hence, the resulting set of equations to be solved is

\begin{eqnarray}
\label{eq:eqgeom}
\begin{array}{lllllll}
X^\prime & = &(\qsubrm{\epsilon}{H}-4)X-\tfrac{1+f_\mathcal{R}-\qsubrm{g}{K}f_\mathcal{R}^\prime}{f_\mathcal{R}^\prime}\qsubrm{\epsilon}{H}Y+\tfrac{1+f_\mathcal{R}+2(1-\qsubrm{g}{K})f_\mathcal{R}^\prime}{f_\mathcal{R}^\prime}\qsubrm{\epsilon}{H}Z,& \quad&{\qsubrm{\Delta^\prime}{m}}  & = &-\qsubrm{g}{K}\qsubrm{\epsilon}{H}\qsubrm{\Theta}{m} +3X,\\
Y^{\prime} & = &-X-2\tfrac{f_\mathcal{R}^\prime}{1+f_\mathcal{R}}Y-\tfrac{1+f_\mathcal{R}-f_\mathcal{R}^\prime}{1+f_\mathcal{R}}Z+\tfrac{1}{1+f_\mathcal{R}}{\qsubrm{\Omega}{m}}{\qsubrm{\Theta}{m}},& \quad&{\qsubrm{\Theta}{m}}^{\prime}& = &-\qsubrm{\epsilon}{H}\qsubrm{\Theta}{m}+ 3Y, \label{eq:eqgeom}\\
Z^\prime& = &X-Y, & & & &
\end{array}
\end{eqnarray}
valid in the conformal Newtonian gauge, where $Z=\phi, Y=\psi$ and $X=\phi^\prime+\psi$. From here, the results in the synchronous gauge can be obtained using the gauge transformation rules.

The last strategy would be to consider only the dark sector fluid variables and geometrical perturbations, eliminating $\qsubrm{\Theta}{m}$ in the equation for $Y$ in (\ref{eq:eqgeom}) with (\ref{eq:Thetam}) and taking the perturbed fluid equations for the dark sector. The resulting set of equations is
\begin{eqnarray}
\label{eq:pfebean}
\begin{array}{lllllll}
X^\prime & = &\tfrac{1+f_\mathcal{R}-\qsubrm{g}{K}f_\mathcal{R}^\prime}{f_\mathcal{R}^\prime (1+f_\mathcal{R})}{\qsubrm{\epsilon}{H}}\chi+(\qsubrm{\epsilon}{H}-4)X+(2-\qsubrm{g}{K})\qsubrm{\epsilon}{H}Z, & \quad\quad&{\qsubrm{{\Theta}}{de}}^{\prime}& = &-3\qsubrm{\Delta}{de}-\qsubrm{\epsilon}{H}\qsubrm{{\Theta}}{de}-2\qsubrm{\Pi}{de}^{\scriptscriptstyle{\mathrm{S}}}-3\qsubrm{\Gamma}{de}, \\
Z^\prime& = &\tfrac{1}{1+f_\mathcal{R}}\chi+X-Z,& \quad\quad& {\qsubrm{\Delta^\prime}{de}} & = &-3\qsubrm{\Delta}{de}-\qsubrm{g}{K}\qsubrm{\epsilon}{H}\qsubrm{{\Theta}}{de}-2\qsubrm{\Pi}{de}^{\scriptscriptstyle{\mathrm{S}}},  \\
\chi^\prime & = & \qsubrm{\Omega}{de}\qsubrm{{\Theta}}{de}+\tfrac{1+f_\mathcal{R}-f_\mathcal{R}^\prime}{1+f_\mathcal{R}}\chi+2f_\mathcal{R}X+f_\mathcal{R}^\prime Z,
& & & &
\end{array}
\end{eqnarray}
with $\qsubrm{\Pi}{de}^{\scriptscriptstyle{\mathrm{S}}}$ and $\qsubrm{\Gamma}{de}$ given in (\ref{eq:eosFL}). In much of the previous work, the linear perturbation in $f(\mathcal{R})$ gravity have been solved with equations analogous to (\ref{eq:pfebean}), with the difference that $\qsubrm{\Delta}{de}$ and $\qsubrm{\hat{\Theta}}{de}$ are replaced with $\qsubrm{\Delta}{m}$ and $\qsubrm{\hat{\Theta}}{m}$, thanks to the field equations (\ref{eq:dgdu}), see \cite{Bean:2006up}. 

We argue that the EoS approach (\ref{eq:pfemde}) provides a clearer set of equations which can be solved and interpreted in an easier way. For instance, the stability of the metric perturbation in the high curvature regime, as discussed in \cite{HSSS}, can be straightforwardly seen in (\ref{eq:pfemde}). When $B\ll1$ (high curvature), from a quick look at the EoS  (\ref{eq:eosPi_S}, \ref{eq:eosGamma}) we can see that

 \be
 \label{eq:design}
 \qsubrm{\Pi}{de}^{\scriptscriptstyle{\mathrm{S}}}=-\tfrac{1}{3\qsubrm{g}{K}\qsubrm{\epsilon}{H}}\mathrm{K}^2\qsubrm{\Delta}{de},\quad\quad\quad\qsubrm{\Gamma}{de}=-\tfrac{2}{3\qsubrm{g}{K}\qsubrm{\epsilon}{H}B}\qsubrm{\Delta}{de}.
 \ee

By considering the second derivative of the perturbed fluid variable $\qsubrm{\Delta}{de}$ from (\ref{eq:pfemde}) in the high curvature regime, one is left with a second order differential equation which can be recast as
 \be
 \label{eq:dpp}
\qsubrm{\Delta}{de}^{\prime\prime}+(3-\tfrac{2}{3\qsubrm{g}{K}\qsubrm{\epsilon}{H}}\mathrm{K}^2)\qsubrm{\Delta}{de}^{\prime}+\tfrac{2}{B}\qsubrm{\Delta}{de}=\tfrac{2}{B}F(\qsubrm{\Delta}{m},\qsubrm{\hat{\Theta}}{m}).
 \ee
Therefore if $B<0$, the perturbed fluid variable $\qsubrm{\Delta}{de}$ does not converge toward the particular solution, $\qsubrm{\Delta}{de}=F(\qsubrm{\Delta}{m},\qsubrm{\hat{\Theta}}{m})$, but diverges exponentially, see \cite{HSSS} for a detailed discussion. 

When a model of $f(\mathcal{R})$ gravity is specified analytically, the EoS for $\qsubrm{\Gamma}{de}$ and $\qsubrm{\Pi}{de}$ (\ref{eq:eosPi_S}, \ref{eq:eosGamma}) are easily obtained from a direct calculation of $f_\mathcal{R}$ and $f_\mathcal{R}^\prime$. For instance, the Hu-Sawicki-Strarobinsky model \cite{Appleby:2009uf,Hu:2007nk,2007JETPL..86..157S},
\be
\label{eq:hssm}
f(\mathcal{R})\equiv-2\Lambda\frac{\left(\frac{\mathcal{R}}{m^{2}}\right)^{2n}}{1+\left(\frac{\mathcal{R}}{m^{2}}\right)^{2n}},
\ee 
commonly used as an alternative to the cosmological constant, leads to the following expressions for $f_\mathcal{R}$ and $f_\mathcal{R}^\prime$,
\bse
\label{eq:hss}
\begin{eqnarray}
f_{\mathcal{R}} & = & 2n\frac{1-\left(\frac{\mathcal{R}}{m^{2}}\right)^{2n}}{1+\left(\frac{\mathcal{R}}{m^{2}}\right)^{2n}}\frac{f(\mathcal{R})}{\mathcal{R}},\\
f_{\mathcal{R}}^{\prime} & = & 2n(2n+1)\frac{\frac{2n-1}{2n+1}+\left(\frac{\mathcal{R}}{m^{2}}\right)^{4n}}{1+2\left(\frac{\mathcal{R}}{m^{2}}\right)^{2n}+\left(\frac{\mathcal{R}}{m^{2}}\right)^{4n}}\frac{\mathcal{R}^{\prime}}{\mathcal{R}}\frac{f(\mathcal{R})}{\mathcal{R}}.\end{eqnarray}\ese

Another interesting case is when the dark energy fluid is dominating ($\qsubrm{\Omega}{m}=0,\,\qsubrm{\Omega}{de}=1$). Then the equation of state parameter reduces to
\be
1+\qsubrm{w}{de}=\tfrac{2}{3}\qsubrm{\epsilon}{H}.
\ee
This case is relevant not only for the present acceleration but also when studying inflationary scenarios based on $f(\mathcal{R})$ modifications to GR. In particular, when the slow-roll conditions are fulfilled, $\qsubrm{\epsilon}{H}\ll1$ and $\qsubrm{\epsilon}{H}^\prime\ll\qsubrm{\epsilon}{H}$, one gets  
\be
\tfrac{d\qsubrm{P}{de}}{d\qsubrm{\rho}{de}}=-1\quad\quad \mathrm{and} \quad\quad \qsubrm{\zeta}{de} =\tfrac{4}{3}\tfrac{\qsubrm{g}{K}-1}{\qsubrm{g}{K}}.
\ee
with $\qsubrm{\epsilon}{H}^\prime=2\qsubrm{\epsilon}{H}^2$ and $\qsubrm{\bar{\epsilon}}{H}=4\qsubrm{\epsilon}{H}$ (see footnote \ref{fn:1}). With no other fluid than the $f(\mathcal{R})$ fluid, the perturbed field equations (\ref{eq:dgdu})  provide very simple relationship between the geometric perturbations and the perturbed fluid variables: $-\tfrac{2}{3}\mathrm{K}^{2}Z  = \qsubrm{\Delta}{de}$ and $2X  =  \qsubrm{\hat{\Theta}}{de}$. For the most popular Starobinsky's proposal \cite{Starobinsky:1980te} for primordial acceleration, 
\be
f(\mathcal{R})=\frac{\mathcal{R}^2}{6M^2},
\ee
one finds $f_\mathcal{R}=\tfrac{2}{3\qsubrm{\epsilon}{H}}-\tfrac{1}{3}$, $f_\mathcal{R}^\prime=-\tfrac{4}{3}$ with $\qsubrm{\epsilon}{H}=\tfrac{M^2}{6H^2}$, as well as $\qsubrm{g}{K}=1+\tfrac{2k^2}{a^2M^2}$. So that the expressions for the anisotropic stress and the entropy perturbation become
\bse
\label{eq:eosGammade}
\bea
\qsubrm{w}{de}\qsubrm{\Pi}{de}^\mathrm{S}  &= & \tfrac{\qsubrm{g}{K}-1}{\qsubrm{g}{K}}\left(\qsubrm{\Delta}{de}+\qsubrm{\epsilon}{H}\qsubrm{\hat{\Theta}}{de} \right),\label{eq:eosPi_Sde}\\
\qsubrm{w}{de}\qsubrm{\Gamma}{de}& = &\tfrac{4}{3\qsubrm{g}{K}}\left\{(\qsubrm{g}{K}+\tfrac{1+\qsubrm{\epsilon}{H}}{\qsubrm{\epsilon}{H}})\qsubrm{\Delta}{de}+\left(\qsubrm{g}{K}-1\right)\qsubrm{\hat{\Theta}}{de}\right\}.
\eea
\ese
These simple expressions can be plugged into the perturbed fluid equations (\ref{eq:pfe}) in order to solve the dynamics of linear perturbations during Starobinsky inflation, in the conformal Newtonian gauge or synchronous gauge. 

\section{Discussion}
The EoS approach for dark sector perturbations has been discussed in details for (i) generalised k-essence theories, where the generic Lagrangian is $\mathcal{L}_{\scriptscriptstyle{\mathrm{de}}}(\phi,\chi)$, with $\chi\equiv-\frac{1}{2}\nabla^\mu \phi\nabla_\mu \phi$, and (ii) for theories in which the dark sector Lagrangian only contains the metric tensor $\mathcal{L}_{\scriptscriptstyle{\mathrm{de}}}(g_{\mu\nu})$, see \cite{Battye:2007aa,2013PhRvD..88h4004B}. In these two cases the gauge invariant equations of state were found to be  
\be
\qsubrm{w}{de}\Pi^{\scriptscriptstyle{\mathrm{S}}}_{\scriptscriptstyle{\mathrm{de}}}=0,\quad\quad\quad\qsubrm{w}{de}\Gamma_{\scriptscriptstyle{\mathrm{de}}}=(\qsubrm{c}{S}^2-\qsubrm{w}{de})\qsubrm{\Delta}{de},
\ee
where $\qsubrm{c}{S}^2$ is the sound speed in the effective dark sector fluid, and 
\be
\qsubrm{w}{de}\Pi^{\scriptscriptstyle{\mathrm{S}}}_{\scriptscriptstyle{\mathrm{de}}}=-\tfrac{3}{2}(\qsubrm{c}{S}^2-\qsubrm{w}{de})\left\{\qsubrm{\Delta}{de}-\qsubrm{\hat{\Theta}}{de}-3(1+\qsubrm{w}{de})Z\right\},\quad\quad\quad\qsubrm{w}{de}\Gamma_{\scriptscriptstyle{\mathrm{de}}}=0,
\ee
respectively.

We have presented the EoS approach to cosmological perturbations in $f(\mathcal{R})$ gravity. After reviewing the formalism for describing the evolution of linear perturbation in $f(\mathcal{R})$ gravity, we have exhibited three equivalent ways to solve their dynamics. In previous work, linear perturbation equations in  $f(\mathcal{R})$ gravity are solved in the geometric picture $(\ref{eq:eqgeom})$. Using the EoS approach (\ref{eq:pfemde}-\ref{eq:pfebean}) appears to have some advantages over the geometrical approach (\ref{eq:eqgeom}) because all the $f(\mathcal R)$ modification can be implemented in the dynamics by simply adding a new fluid species at the perturbed level, rather than modifying the whole set of equation for the geometrical variables. 

The main results of this paper are the equations of state  for $\qsubrm{\Gamma}{de}$ and $\qsubrm{\Pi}{de}$ (\ref{eq:dudefvt}, \ref{eq:eosPi_S}, \ref{eq:eosGamma}, \ref{eq:eosFL}). In these expressions the entropy perturbation and the anisotropic stresses are specified either in terms of the perturbed fluid variables of the dark sector and standard matter fluid, $\Pi^{\scriptscriptstyle{\mathrm{S}}}_{\scriptscriptstyle{\mathrm{de}}}=\Pi^{\scriptscriptstyle{\mathrm{S}}}_{\scriptscriptstyle{\mathrm{de}}}(\qsubrm{\Delta}{de},\qsubrm{\hat{\Theta}}{de},\qsubrm{\Delta}{m},\qsubrm{\hat{\Theta}}{m})$ and $\Gamma_{\scriptscriptstyle{\mathrm{de}}}=\Gamma_{\scriptscriptstyle{\mathrm{de}}}(\qsubrm{\Delta}{de},\qsubrm{\hat{\Theta}}{de},\qsubrm{\Delta}{m},\qsubrm{\hat{\Theta}}{m})$, or the perturbed fluid variables of the dark sector and the geometrical perturbations, $\Pi^{\scriptscriptstyle{\mathrm{S}}}_{\scriptscriptstyle{\mathrm{de}}}=\Pi^{\scriptscriptstyle{\mathrm{S}}}_{\scriptscriptstyle{\mathrm{de}}}(Y,Z)$ and $\Gamma_{\scriptscriptstyle{\mathrm{de}}}=\Gamma_{\scriptscriptstyle{\mathrm{de}}}(\qsubrm{\Delta}{de},\qsubrm{\hat{\Theta}}{de},X,Y,Z)$, thanks to the field equations (\ref{eq:dgdu}). An important point is the extra degree of freedom, $\hat{\chi}$, induced by a non-trivial  $f(\mathcal{R})$ modification to GR, is absent of these expressions. The elimination of this internal degree of freedom is the essence of the procedure. 

In order to illustrate the EoS formalism we have presented the EoS in the scalar sector for three different cases: (i) the designer $f(\mathcal{R})$ in the high curvature regime (\ref{eq:design}), (ii) the analytical Hu-Sawicki-Starobinsky model for dark energy (\ref{eq:hss}; to be plugged into \ref{eq:eosPi_S}, \ref{eq:eosGamma});  and (iii) the Starobinsky proposal for inflation (\ref{eq:eosGammade}).

\section*{Acknowledgements}
JAP  is supported by the STFC Consolidated Grant ST/J000426/1. BB is supported by a grant from ENS de Lyon.

\clearpage
\appendix
\section{Gauge transformation rules}
\label{append_gauge}

Any expression written in the conformal Newtonian gauge can by translated
into the synchronous gauge and vice versa. The relationship between
both gauges can be seen as an infinitesimal coordinate transformation,
with a specific four vector $d^{\mu}$ characterizing the change of
coordinate \cite{Ma:1995ey}. The time-like component of $d^{\mu}$ is
\begin{equation}
{d^0}=\frac{a^2\dot{h}_{\parallel}}{2k^{2}}.
\end{equation}
It is related to $T$ introduced in section \ref{sec:gif} by $T=Hd^0$. The dot denotes derivative with respect to coordinate time. The transformation rules for the metric perturbations, from the conformal Newtonian gauge to the synchronous gauge are
\begin{eqnarray}
\psi & = & \dot{{d^0}},\nonumber \\
\phi & = & \eta-H{d^0},\nonumber \\
\dot{\phi}+H\psi & = & \dot{\eta}-\dot{H}{d^0},\label{eq:tr geo}\\
\ddot{\phi}+H\dot{\psi}+2\dot{H}\psi & = & \ddot{\eta}-\ddot{H}{d^0}.\nonumber 
\end{eqnarray}
For the fluid perturbations, the transformation rules are
\begin{eqnarray}
\delta_{c} & = & \delta_{s}+\frac{\dot{\rho}}{\rho}d^0,\nonumber \\
\theta_{c} & = & \theta_{s}+{d^0},\nonumber \\
\delta P_{c} & = & \delta P_{s}+\dot{P}d^0,\\
\qsuprm{\Pi_{c}}{S,V,T} & = & \qsuprm{\Pi_{s}}{S,V,T},\nonumber 
\end{eqnarray}
where the subscripts `c' and `s' hold for conformal Newtonian gauge and synchronous gauge respectively. For $\chi$ and its time derivatives the transformation rules are obtained from its definition in terms of the first order perturbation of the Ricci scalar (\ref{eq:chidef})
\begin{eqnarray}
\mathcal{\chi}_{c} & = & \chi_{s}+\dot{f_{\mathcal{R}}}{d^0},\nonumber \\
\dot{\chi}_{c}- \dot{f_{\mathcal{R}}}\psi & = & \dot{\chi}_{s}+\ddot{f_{\mathcal{R}}}{d^0},\label{eq:tr chi}\\
\ddot{\chi}_{c}-\dot{f_{\mathcal{R}}}\dot{\psi}-2\ddot{f_{\mathcal{R}}}\psi & = & \ddot{\chi}_{s}+\dddot{f_{\mathcal{R}}}{d^0}.\nonumber 
\end{eqnarray}
Writing these relations with $T$  instead of $d^0$ and the `prime' derivative instead of the `dot' derivative leads to the definition of the gauge invariant notations of section \ref{sec:gif}. 

\bibliographystyle{JHEP}
\bibliography{refs}

\end{document}

%% file: defs.tex
\newcommand{\dd}[0]{\textrm{d}}
\newcommand{\defn}[0]{\equiv}

\newcommand{\qsubrm}[2]{{#1}_{\scriptscriptstyle{\textrm{#2}}}}

\newcommand{\qsuprm}[2]{{#1}^{\scriptscriptstyle\textrm{#2}}}

\newcommand{\pis}[0]{ {\Pi}^{\scriptscriptstyle\rm{S}}}

\newcommand{\rbm}[1]{{\bf{#1}}}
\newcommand{\ci}[0]{\textrm{i}}

\def\be{\begin{equation}}
\def\ee{\end{equation}}
\def\bea{\begin{eqnarray}}
\def\eea{\end{eqnarray}}
\def\bse{\begin{subequations}}
\def\ese{\end{subequations}}

\let\oldsqrt\sqrt
\def\sqrt{\mathpalette\DHLhksqrt}
\def\DHLhksqrt#1#2{%
\setbox0=\hbox{$#1\oldsqrt{#2\,}$}\dimen0=\ht0
\advance\dimen0-0.2\ht0
\setbox2=\hbox{\vrule height\ht0 depth -\dimen0}%
{\box0\lower0.4pt\box2}}